\documentclass[12pt]{iopart}

\usepackage{iopams}
\usepackage{amssymb}  

\begin{document}

\title[Restricted Incompressible Navier--Stokes Equations]{Exact Solutions for Restricted Incompressible Navier--Stokes Equations with Dirichlet Boundary Conditions.}

\author{Manuel Garc\'ia-Casado}

\address{Energy Department. T\"UV S\"UD Iberia. Tres Cantos. 28760 Madrid. Spain}
\ead{manuel.garciacasado@gmail.com}
\vspace{10pt}
\begin{indented}
\item[]February 2019
\end{indented}

\begin{abstract}
This paper exposes how to obtain a relation that have to be hold for all free-divergence velocity fields that evolve according to Navier--Stokes equations. However, checking the violation of this relation requires a huge computational effort. To circumvent this problem it is proposed an additional ansatz to free-divergence Navier--Stokes fields. This makes available six degrees of freedom which can be tuned. When they are tuned adequately, it is possible to find finite $L^2$ norms of the velocity field for volumes of $\mathbb{R}^3$ and for $t\in[t_0,\infty)$. In particular, the kinetic energy of the system is bounded when the field components $u_i$ are class $C^3$ functions on $\mathbb{R}^3\times[t_0,\infty)$ that hold Dirichlet boundary conditions. This additional relation lets us conclude that Navier--Stokes equations with no-slip boundary conditions have not unique solution. Moreover, under a given external force the kinetic energy can be computed exactly as a funtion of time.
\end{abstract}

%\submitto{\JPCO}
 This is an author-created, un-copyedited version of an article published in Journal of  Physics Communications. IOP Publishing Ltd is not responsible for any errors or omissions in this version of the manuscript or any version derived from it. This article is published under a CC BY licence. The Version of Record is available online at https://doi.org/10.1088/2399-6528/ab3837
%\maketitle

\section{Introduction}

The evolution of a system is represented by means of magnitudes that change over time. Typically, the dynamical system is defined by differential equations of time functions. However, this definition becomes inconsistent when time integration of those equations does not guarantee that magnitudes are finite all the time. In systems like incompressible Navier--Stokes equations, it is imperative to find additional restrictions (or equations) to avoid that situation in which the functions become infinite at finite time \cite{RefJ1},  \cite{RefJ2}. Moreover, it is necessary in these advective velocity fields to assure that the kinetic energy remains bounded, at least for a short time \cite{RefB2}. Many methods for these equations to find weak solutions have been developed \cite{RefJ3}-\cite{RefJ5}, but still is not clear that such systems have unique solutions. There was proved in \cite{RefB10}, \cite{RefJ5.01} that if there exists a classical solution in a connected subset of $\mathbb{R}^3\times [t_0,T]$ then it is also a Leray-Hopf weak solution \cite{RefJ3}, \cite{RefJ8},\cite{RefJ9}. It is also proved that if there exists a Leray-Hopf weak solutions in $\mathbb{R}^3\times [t_0,T]$, it is a unique solution. Conversely, if there is a uniquely weak solution $\bi{u}$ with partial derivatives $\partial_i \partial_j u_k$ belonging to $L^2(\mathbb{R}^3\times [t_0,T])$, then this one is also a classical solution for the Navier--Stokes equation. However, it has been proposed recently that Navier--Stokes equations have not unique weak solutions \cite{RefJ5.1}. The present paper is in the line of this recent paper.
In the first part of the present paper, we expose how to find a relation for the velocity field components and derivatives. This relation is an inequality that involves second derivative in time of the sphere area. To see where this relationship comes from, we expose what conditions are needed for surface area of a volume to grow over time. The dynamic of surfaces has gained attention in last decades. Fluid surfaces dynamics are applied to interfaces separating fluid phases \cite{RefJ5.12}, thin films of liquid \cite{RefJ5.13}, liquid layers with surfactants \cite{RefJ5.14}, even to biofilms in porous media \cite{RefJ5.15}. These studies as the relation obtained here come from the transport theorem for surfaces. In our case, the surface encloses a volume. When it is a ball, it is necessary to compute the second time derivative of the surface area and to particularize this result to the sphere. These inequalities have to be hold for all possible surface balls in the domain of the velocity field in $\mathbb{R}^3\times [t_0,\infty)$. However, this requires a huge computational effort. If the second time derivative of the area is applied to velocity fields that hold Navier--Stokes equations, we could realize that an additional relationship is needed between spatial second derivative of pressure and spatial derivatives of velocity field components. To circumvent the computational problem, in the second part of the present paper it is suggested an equation that, when it holds, guarantees that volume integral of a velocity norm is finite at every time lapse under suitable boundary conditions. The restriction exposed in the second part of this paper is a matrix relation between spatial partial derivatives of velocity and pressure. This type of relations have been treated earlier \cite{RefJ5.16},\cite{RefJ5.17}, but the analysis in these works resides in statistical properties. These types of ansatz are common in dynamic systems \cite{RefJ5.2}-\cite{RefJ5.4} since they allow to observe the problem under different points of view. Then, if the restriction to Navier--Stokes equations exposed here were maintained, the kinetic energy, the volume integral of the velocity field norm can be computed exactly. However, if we do not imposes this restriction, we have several solutions at the same time for no-slip boundary. To find a numerical solution to Navier--Stokes equations with no-slip boundary is a long-standing problem in numerical simulations \cite{RefJ5.5}. The incompressibility is approximated by artificial compressibility. This artefact is widely used until now \cite{RefB5.6}. Recently, there had appeared computational methods to make the energy stable by mean of small variable in density and viscosity \cite{RefJ5.7}. But these are approximations too. Here, the system of equations is solved analytically in the sense that kinetic energy can be worked out as time function without approximations. 

Then, the first issue is to obtain a transport theorem for surfaces. This theorem is not new but helps us to fix the notation. Second, we will show a differential relation of velocity and pressure that is a generalization of the Poisson equation for the pressure. This relation gives us a bounding for the infinitesimal strain tensor of the fluid. Finally, if the velocity is null outside the considered volume, those assumptions allow us to obtain an upper bound to the volume integral of quadratic sum of velocity field components. In this way, we give an example in which the kinetic energy can be computed exactly.

\section{Transport theorem for surfaces}
For technical reasons, Reynold's transport theorem  \cite{RefB3}-\cite{RefB8} is a very useful tool since it allows us introduce the time derivative of a dynamic volume integral inside the integrand of a static integral. The change in the volume shape pass to the integrand. The same can be done with surface integrals. The integrating surface is moving and changing its shape over time. Then, the transformation from time derivative of a surface integral to a surface integral is not immediate task since, in general, time derivative and dynamic integral does not permute. To pass the time derivative inside the integrand of the moving surface require some effort. For this purpose, the velocity field considered here is defined as the vector-valued function \(\bi{u}:\mathbb{R}^3  \times  [t_0, \infty) \longrightarrow \mathbb{R}^3\) with components \(u_i, i\in \{1,2,3\}\); and \(t_0\) is the initial time. Moreover, \(\Omega\subset \mathbb{R}^3\) is a volume dragged by the velocity field. Its boundary is the closed surface \(\Sigma \equiv \partial \Omega\). Formally, let \(\bi{x}\in \Omega \cup \Sigma \) and let \(\phi_{t}\) denote the invertible mapping  \(\bi{x}\longmapsto \bi{\phi}(\bi{x},t)\in \mathbb{R}^3\) which  can be viewed as the flow with properties \(\bi{\phi}(\bi{\phi}\left(\bi{x},s\right),t)=\bi{\phi}\left(\bi{x},s+t\right)\) and \(\bi{\phi}(\bi{x},t_0)=\bi{x}\). So, \(\phi_{t}\) is the mapping that takes the volume \(\Omega\) at time \(t_{0}\) to the volume \(\Omega_{t}\) at time \(t\), and hence, it also takes the surface \(\Sigma\) at time \(t_{0}\) to the surface \(\Sigma_{t}\) at time \(t\). In this way, the velocity is given by \(\bi{u}\left(\bi{\phi}\left(\bi{x},t\right),t\right)\equiv \frac{\partial \bi{\phi}\left(\bi{x},t\right) }{\partial t}\). Moreover,  \(\bi{x}\) can be considered as the parametrization of the surface \(\Sigma\) that takes  \(\left(\alpha, \beta\right)\in [0,1]\times[0,1]\subset \mathbb{R}^2\) to \(\bi{x}\left(\alpha, \beta\right)\in \Sigma\subset \mathbb{R}^3\). This allows to define the unit normal vector \(\bi{n}\) to the surface \(\Sigma_t\), with components \(n_i\), as 
\begin{eqnarray}
\bi{n}=\frac{\partial_{\alpha} \bi{\phi}\times\partial_{\beta} \bi{\phi}}{\Vert \partial_{\alpha} \bi{\phi}\times\partial_{\beta} \bi{\phi}\Vert}
\label{eq section 2 001} 
\end{eqnarray}
where  \(\partial_{\alpha} \bi{\phi}=\frac{\partial\bi{x}}{\partial \alpha}\cdot\bi{\nabla}\bi{\phi}\left(\bi{x},t\right)\) , \(\partial_{\beta} \bi{\phi}=\frac{\partial\bi{x}}{\partial \beta}\cdot\bi{\nabla}\bi{\phi}\left(\bi{x},t\right)\); and \(\Vert \cdot \Vert\) is the Euclidean norm. Moreover, the material derivative of \(f\left(\bi{x},t\right) \in \mathbb{R}\) is defined as \(\frac{Df}{Dt} \equiv \frac{\partial f}{\partial t}+\bi{u}\cdot \bi{\nabla} f\). With these definitions, the theorem can be stated as follows.
\subsubsection*{\bf{Theorem 1.}}	Let \(\bi{u}\) be a differentiable velocity field as defined above, \(f\left(\bi{x},t\right) \in \mathbb{R}\) be a smooth function and \(\Omega_t\) be a Lebesgue measurable domain with smooth boundary. Then, the time derivative of the surface integral over \(\Sigma_t\) of the function transported by the field is
	\begin{eqnarray}
	\frac{d}{dt}\int_{\Sigma_{t}}f d^{2}x=\int_{\Sigma_{t}}\left[\frac{D f}{Dt} + f \bi{\nabla} \cdot \bi{u}-f \bi{n}\cdot\left(\bi{n}\cdot \bi{\nabla} \bi{u}\right)\right]d^{2}x.
	\label{eq section 2 002}  
	\end{eqnarray}
\subsubsection*{Proof.}The moving surface \(\Sigma_{t}\) can be parametrized by \(\alpha\in[0,1]\) and \(\beta\in[0,1]\). So the equation \eref{eq section 2 002} can be rewritten as 
\begin{eqnarray}
\frac{d}{dt}\int_{\Sigma_{t}}f\left(\bi{x},t\right) d^{2}x \nonumber\\
=\frac{d}{dt}\int^{1}_{0}\int^{1}_{0}f\left(\bi{\phi}\left(\bi{x}(\alpha,\beta),t\right),t\right)\Vert\partial_{\alpha}\bi{\phi}\times\partial_{\beta}\bi{\phi}\Vert\left(\bi{x}(\alpha,\beta),t\right) d\alpha d\beta.
\label{eq section 2 003} 
\end{eqnarray}
Now, the integration limits do not depend on time and the time derivative passes into the integrand. Then,
\begin{eqnarray}
\frac{d}{dt}\int_{\Sigma_{t}}f\left(\bi{x},t\right) d^{2}x \nonumber\\
=\int^{1}_{0}\int^{1}_{0}\left\{\frac{d}{dt}\left[ f\left(\bi{\phi}\left(\bi{x}(\alpha,\beta),t\right),t\right)\right]\Vert\partial_{\alpha}\bi{\phi}\times\partial_{\beta}\bi{\phi}\Vert\left(\bi{x}(\alpha,\beta),t\right) \right. \nonumber\\ \left. +f\left(\bi{\phi}\left(\bi{x}(\alpha,\beta),t\right),t\right)\frac{d}{dt}\left[\Vert\partial_{\alpha}\bi{\phi}\times\partial_{\beta}\bi{\phi}\Vert\left(\bi{x}(\alpha,\beta),t\right)\right]\right\} d\alpha d\beta.
\label{eq section 2 004} 
\end{eqnarray}
The chain rule can be applied to both time derivatives of right hand side of \eref{eq section 2 004}. The first one is
\begin{eqnarray}
\frac{d}{dt}\left[ f\left(\bi{\phi}\left(\bi{x}(\alpha,\beta),t\right),t\right)\right] \nonumber\\
=\frac{\partial}{\partial t}f\left(\bi{\phi}\left(\bi{x}(\alpha,\beta),t\right),t\right)+\frac{\partial}{\partial t} \bi{\phi}\left(\bi{x}(\alpha,\beta),t\right)\cdot \bi{\nabla_\phi}f\left(\bi{\phi}\left(\bi{x}(\alpha,\beta),t\right),t\right)
\label{eq section 2 005} 
\end{eqnarray}
where \(\bi{\nabla_{\phi}}\) is the gradient built from partial derivatives with respect of \(\bi{\phi}\) components. The second time derivative of right hand side of \eref{eq section 2 004}  is 
\begin{eqnarray}
\frac{d}{dt}\Vert\partial_{\alpha}\bi{\phi}\times\partial_{\beta}\bi{\phi}\Vert \nonumber\\
=\frac{\left(\partial_{\alpha}\bi{\phi}\times\partial_{\beta}\bi{\phi}\right)}{\Vert\partial_{\alpha}\bi{\phi}\times\partial_{\beta}\bi{\phi}\Vert}\cdot\left[\left(\partial_{\alpha}\bi{\phi}\cdot\bi{\nabla_{\phi}}\frac{\partial \bi{\phi}}{\partial t}\right)\times\partial_{\beta}\bi{\phi}+\partial_{\alpha}\bi{\phi}\times\left(\partial_{\beta}\bi{\phi}\cdot\bi{\nabla_{\phi}}\frac{\partial \bi{\phi}}{\partial t}\right)\right]
\label{eq section 2 006} 
\end{eqnarray}
where the functions arguments are omitted for clarity. A little more algebra transforms this relation into
\begin{eqnarray}
\frac{d}{dt}\Vert\partial_{\alpha}\bi{\phi}\times\partial_{\beta}\bi{\phi}\Vert \nonumber\\
=\Vert \partial_{\alpha}\bi{\phi}\times\partial_{\beta}\bi{\phi}\Vert \left[\bi{\nabla_{\phi}}\cdot\frac{\partial \bi{\phi}}{\partial t}\right]-\left[\left(\partial_{\alpha}\bi{\phi}\times\partial_{\beta}\bi{\phi}\right)\cdot \bi{\nabla_{\phi}}\frac{\partial \bi{\phi}}{\partial t}\right] \cdot\frac{\left(\partial_{\alpha}\bi{\phi}\times\partial_{\beta}\bi{\phi}\right)}{\Vert\partial_{\alpha}\bi{\phi}\times\partial_{\beta}\bi{\phi}\Vert}.
\label{eq section 2 007} 
\end{eqnarray}
Plugging these results in \eref{eq section 2 004}, and taking into account the definition of the normal vector to the surface \eref{eq section 2 001} and that \(\bi{u}\left(\bi{\phi}\left(\bi{x},t\right)\,t\right) = \frac{\partial \bi{\phi}\left(\bi{x},t\right) }{\partial t}\), it is found that 
\begin{eqnarray}
\frac{d}{dt}\int_{\Sigma_{t}} f\left(\bi{x},t\right) d^2 x =\int^{1}_{0}\int^{1}_{0}\left[\frac{\partial f}{\partial t}+ \bi{u}\cdot \bi{\nabla_{\phi}} f  \right. \nonumber\\  + f\left\{\bi{\nabla_{\phi}}\cdot\bi{u} - \left(\bi{n} \cdot \bi{\nabla_{\phi} u}\right) \cdot \bi{n}\right\} \bigg]\Vert\partial_{\alpha}\bi{\phi}\times\partial_{\beta}\bi{\phi}\Vert d\alpha d\beta  
\label{eq section 2 008} 
\end{eqnarray}
Finally, undoing the surface parametrization, this last relation gives us the theorem result
\begin{eqnarray}
\frac{d}{dt}\int_{\Sigma_{t}} f\left(\bi{x},t\right) d^2 x =\int_{\Sigma_{t}}\left[\frac{D f}{D t}+ f\left\{\bi{\nabla}\cdot\bi{u} - \left(\bi{n} \cdot \bi{\nabla u}\right) \cdot \bi{n}\right\} \right] d^2 x.
\label{eq section 2 009} 
\end{eqnarray}
\hfill\ensuremath{\square}

Equation (\ref{eq section 2 002}) is similar to the transport theorem for moving surfaces of volumes \cite{RefJ6}, \cite{RefJ7}, which is usually written in terms of both, normal velocity and curvature of the surface. However, in this case, the term corresponding to the boundary of the surface is missing since it is a closed one and, hence, it has not boundary. Perhaps, the normal vector $\bi{n}$ inside the integrand could be confusing since it depends on the surface choice, but notice that we can rewrite the surface integral as the identity
\begin{eqnarray}
\int_{\Sigma_{t}} d^2 x \equiv\int_{\Sigma_{t}}\bi{n}\cdot\bi{n}d^2 x\equiv \int_{\Sigma_{t}}\delta_{ij}n_in_jd^2 x,
\label{eq section 2 010} 
\end{eqnarray}
where we have used Einstein notation for summation on repeated indexes and $\delta_{ij}$ is the Kronecker delta. With this notation, the formula of the theorem can be rewritten as
\begin{eqnarray}
\frac{d}{dt}\int_{\Sigma_{t}} f d^2 x =\int_{\Sigma_{t}}\left[\left(\frac{D f}{D t}+ f\partial_ku_k\right)\delta_{ij}- f\partial_iu_j \right] n_i n_j d^2 x.
\label{eq section 2 011} 
\end{eqnarray}
Then, it is easier to compute  the second derivative of a surface integral of the function $f$. But here, we show that the second time derivative of a surface is useful to obtain an equation that we will use later. So, from  (\ref{eq section 2 002}) we have that
\begin{eqnarray}
\frac{d^2}{dt^2}\int_{\Sigma_{t}} fd^2 x =\frac{d}{dt}\int_{\Sigma_{t}}\left[\left(\frac{D f}{D t}+ f\partial_ku_k\right)\delta_{ij}- f\partial_iu_j \right] n_i n_j d^2 x
\label{eq section 2 012} 
\end{eqnarray}
and then
\begin{eqnarray}
\frac{d^2}{dt^2}\int_{\Sigma_{t}}f d^{2}x=\int_{\Sigma_{t}}\left\{\frac{D }{D t}\left[\left(\frac{D f}{D t}+ f\partial_ku_k\right)\delta_{ij}- f\partial_iu_j \right] n_i n_j
\right.\nonumber\\
+\left.\left[\left(\frac{D f}{D t}+ f\partial_ku_k\right)\delta_{ij}- f\partial_iu_j \right] \frac{Dn_i}{D t} n_j\right.\nonumber\\
+\left.\left[\left(\frac{D f}{D t}+ f\partial_ku_k\right)\delta_{ij}- f\partial_iu_j \right] n_i \frac{D n_j}{D t}\right.\nonumber\\
+\left.\left[\left(\frac{D f}{D t}+ f\partial_ku_k\right)\delta_{ij}- f\partial_iu_j \right] n_i n_j\left[\partial_mu_m\delta_{kl}-\partial_ku_l\right]n_kn_l \right\}d^2x.
\label{eq section 2 013} 
\end{eqnarray}
To simplify this equation we use 
\begin{eqnarray}
\frac{Dn_i}{Dt}
=\frac{dn_i}{dt}=-n_l\partial_lu_i+n_in_ln_k\partial_lu_k,
\label{eq section 2 014} 
\end{eqnarray}
that is deduced from relation (\ref{eq section 2 006}), and $n_in_i=1$ to give
\begin{eqnarray}
\frac{d^2}{dt^2}\int_{\Sigma_{t}}f d^{2}x=\int_{\Sigma_{t}}\left\{\frac{D }{D t}\left[\left(\frac{D f}{D t}+ f\partial_ku_k\right)\delta_{ij}- f\partial_iu_j \right] n_i n_j
\right.\nonumber\\
+\left.f\left[n_jn_l\partial_lu_i\partial_iu_j +n_in_l\partial_lu_j\partial_iu_j-2(n_in_j\partial_iu_j)^2\right]\right.\nonumber\\
+\left.\left[\left(\frac{D f}{D t}+ f\partial_ku_k\right)\delta_{ij}- f\partial_iu_j \right] n_i n_j\left[\partial_mu_m\delta_{kl}-\partial_ku_l\right]n_kn_l \right\}d^2x.
\label{eq section 2 015} 
\end{eqnarray}
This raw equation gives de second time derivative of the surface integral of a function that is dragged by a velocity field. When this function is the density $\rho(\bi{x},t)$ of the fluid, (\ref{eq section 2 015}) can be simplified to
\begin{eqnarray}
\frac{d^2}{dt^2}\int_{\Sigma_{t}}\rho d^{2}x=\int_{\Sigma_{t}}\rho\left\{-\frac{D }{D t}\left( \partial_iu_j \right) n_i n_j
+n_jn_l\partial_lu_i\partial_iu_j \right.\nonumber\\
\left.+n_in_l\partial_lu_j\partial_iu_j-(n_in_j\partial_iu_j)^2 \right\}d^2x,
\label{eq section 2 016} 
\end{eqnarray}
using the continuity equation
\begin{eqnarray}
\frac{D \rho}{D t}+ \rho\partial_ku_k=0.
\label{eq section 2 016.1} 
\end{eqnarray}
Moreover we have the identity
\begin{eqnarray}
n_in_l\partial_lu_j\partial_iu_j-(n_in_j\partial_iu_j)^2\nonumber\\
=\partial_ju_l\partial_iu_m n_in_jn_kn_n(\delta_{lm}\delta_{kn}-\delta_{ln}\delta_{km})\nonumber\\
=(\epsilon_{alk}n_k n_j\partial_ju_l)(\epsilon_{abc}n_cn_i\partial_iu_b),
\label{eq section 2 016.2} 
\end{eqnarray}
where $\epsilon_{abc}$ with $a,b,c\in\{1,2,3\}$ is the Levi-Civita tensor, so
\begin{eqnarray}
\frac{d^2}{dt^2}\int_{\Sigma_{t}}\rho d^{2}x=\int_{\Sigma_{t}}\rho\left\{-\frac{D }{D t}\left( \partial_iu_j \right) n_i n_j
+n_jn_l\partial_lu_i\partial_iu_j \right.\nonumber\\
\left.+(\epsilon_{alk}n_k n_j\partial_ju_l)(\epsilon_{abc}n_cn_i\partial_iu_b)\right\}d^2x.
\label{eq section 2 016.3} 
\end{eqnarray}

Now that we know the rate of change of the surface integral of a magnitude with time, we would like to know whether the area of the surface grows, diminishes or remains constant with time when the volume does not change. A particular case is the sphere, the surface of a ball. One of the sphere properties is that it has the least area that encloses a volume \cite{RefB9}, \cite{RefJ10}. So, the area of the sphere only can increase or be the same few time later. This means that the area is a convex function of time near the minimum. The next theorem depicts this situation.
\subsubsection*{\bf{Theorem 2.}} Let $\bi{u}$ be a class $C^3$ velocity field as defined above. Let $\mathbb{S}^{3}\subset\mathbb{R}^{3}$ be balls with boundaries $\mathbb{S}^{2}\subset\mathbb{R}^{3}$. Also, there exists only one region $\Omega_t\subset \mathbb{R}^{3}$ for $t \neq t_{0}$ such as $\Omega_t\rightarrow S^{3}$ when $t\rightarrow t_{0}$, where $S^3\in\mathbb{S}^{3}$. For every $t$, if the velocity field holds the incompressibility statement, $\bi{\nabla}\cdot\bi{u}=0$, then
\begin{eqnarray}
\int_{S^2}\left\{-\frac{D }{D t}\left( \partial_iu_j \right) n_i n_j	+n_jn_l\partial_lu_i\partial_iu_j \right.\nonumber\\
\left.+(\epsilon_{alk}n_k n_j\partial_ju_l)(\epsilon_{abc}n_cn_i\partial_iu_b)\right\}d^2x\geq 0
\label{eq section 2 018} 
\end{eqnarray}
where $S^2$ is the boundary of $S^3$.
\subsubsection*{Proof.}Taking into account the very well known isoperimetric inequality for three dimensions \cite{RefB9},\cite{RefJ10}, we have
\begin{eqnarray}
\int_{\Sigma_t}d^{2}x \geq 3 \left(\frac{4}{3}\pi \right)^{\frac{1}{3}}\left[\int_{\Omega_t}d^{3}x\right]^{\frac{2}{3}},
\label{eq section 2 019} 
\end{eqnarray}
where the equality holds for the ball $S^{3}$.
We subtract the area of $S^{2}$ on both sides,
\begin{eqnarray}
\int_{\Sigma_t}d^{2}x-\int_{S^{2}}d^{2}x \geq 3 \left(\frac{4}{3}\pi \right)^{\frac{1}{3}}\left[\int_{\Omega_t}d^{3}x\right]^{\frac{2}{3}}-\int_{S^{2}}d^{2}x\nonumber\\
\geq
3 \left(\frac{4}{3}\pi \right)^{\frac{1}{3}}\left\{\left[\int_{\Omega_t}d^{3}x\right]^{\frac{2}{3}}-\left[\int_{S^{3}}d^{3}x\right]^{\frac{2}{3}}\right\}.
\label{eq section 2 020} 
\end{eqnarray}
Due to the incompressibility of the fluid, $S^{3}$ and $\Omega$ have the same volume. The right hand side of (\ref{eq section 2 020}) then vanishes
\begin{eqnarray}
\int_{\Sigma_t}d^{2}x-\int_{S^{2}}d^{2}x \geq 0.
\label{eq section 2 021} 
\end{eqnarray}
In addition, the area time derivative is given by (\ref{eq section 2 002}), with $f= 1$ and $\partial_{i}u_{i}= 0$,
\begin{eqnarray}
\left[\frac{d}{dt}\int_{\partial\Omega_t}d^2 x\right]\left(t_0\right)=-\int^{\pi}_{0}\int^{2\pi}_{0}\partial_{r}u_{r} r^{2} \sin \theta d\theta d\phi\nonumber\\=-\partial_{r}\left[\int^{\pi}_{0}\int^{2\pi}_{0}u_{r} r^{2} \sin \theta d\theta d\phi\right]+\frac{2}{r}\int^{\pi}_{0}\int^{2\pi}_{0}u_{r} r^{2} \sin \theta d\theta d\phi\nonumber\\=-\partial_{r}\left[\int_{S^3}\partial_{i}u_{i} d^3x\right]+\frac{2}{r}\int_{S^3}\partial_{i}u_{i} d^3x=0.
\label{eq section 2 022} 
\end{eqnarray}
So the area of a sphere reaches its minimum at time $t=t_0$ in a incompressible velocity field. This property together with (\ref{eq section 2 021}) means that the area is a local convex function of time in a range close to $t_{0}$.
Therefore, the second time derivative of this function at $t_{0}$ holds
\begin{eqnarray}
\left[\frac{d^{2}}{dt^{2}}\int_{\Sigma_t}d^2 x\right]\left(t_0\right) \geq 0.
\label{eq section 2 023} 
\end{eqnarray}
The second time derivative of the area can be computed applying (\ref{eq section 2 016.3}) for $\rho=1$, giving rise to
\begin{eqnarray}
\left[\frac{d^{2}}{dt^{2}}\int_{\Sigma_t}d^2 x\right]\left(t_0\right)=\left[\int_{\Sigma_t}\left\{-\frac{D }{D t}\left( \partial_iu_j \right) n_i n_j
+n_jn_l\partial_lu_i\partial_iu_j\right.\right.\nonumber\\
\left.\left.+(\epsilon_{olk}n_k n_j\partial_ju_l)(\epsilon_{omn}n_nn_i\partial_iu_m)\right\}d^2x\right]\left(t_0\right) \nonumber\\
=\int_{S^2}\left\{-\frac{D }{D t}\left( \partial_iu_j \right) n_i n_j
+n_jn_l\partial_lu_i\partial_iu_j\right.\nonumber\\
\left.+(\epsilon_{alk}n_k n_j\partial_ju_l)(\epsilon_{abc}n_cn_i\partial_iu_b)\right\}d^2x\geq 0. 		
\label{eq section 2 024} 
\end{eqnarray}
\hfill\ensuremath{\square}

In this equation we see that, at every time, for every spherical surface, there exist a volume, which is a function of time, that converges to the ball. Then (\ref{eq section 2 018}) is held at every instant of time.
For Theorem 2, given that we have a surface integral, it does not matter what velocity distribution is inside the ball but just on its surface. Therefore, this theorem asserts that if there exist at least a sphere in the domain of the incompressible velocity field that violates (\ref{eq section 2 018}), time evolution  for that velocity field is forbidden. The next result applies this last theorem to incompressible Navier--Stokes fluids. 
\subsubsection*{\bf{Theorem 3.}} Let $p$ be the pressure defined as the class $C^2$ function $p:\mathbb{R}^3  \times  [t_0, \infty) \longrightarrow \mathbb{R}$. Let $\bi{u}$ be an incompressible class $C^3$ velocity field as defined above, $\bi{\nabla}\cdot \bi{u}=0$, which evolves in time according to the Navier--Stokes equations 
	\begin{eqnarray}
	\partial_{t}\bi{u}+\bi{u}\cdot\bi{\nabla}\bi{u}=\nu \triangle\bi{u}-\bi{\nabla} p.
	\label{eq section 2 025} 
	\end{eqnarray}
	Here, the density is $\rho=1$ and $\nu$ is the viscosity. Then, at every time $t$, for every spherical region $S^3\in\mathbb{S}^{3}\subset\mathbb{R}^{3}$ with boundary $S^{2}$, we have
	\begin{eqnarray}
	\int_{S^2}\left\{\left(2\partial_ju_k\partial_ku_i+\partial_i\partial_j p-\nu\partial_k\partial_k \partial_i u_j\right) n_i n_j \right.\nonumber\\
	\left.+(\epsilon_{alk}n_k n_j\partial_ju_l)(\epsilon_{abc}n_cn_i\partial_iu_b)\right\}d^2x\geq 0,\label{eq section 2 026} 
	\end{eqnarray}
	where we have used Einstein notation for repeated indices.
\subsubsection*{Proof.} Substitution of relation (\ref{eq section 2 025}) on (\ref{eq section 2 018}) gives rise to (\ref{eq section 2 026}).\hfill\ensuremath{\square}

This theorem establishes that if we find at least a sphere for which the incompressible velocity field does not hold (\ref{eq section 2 026}), that field can not evolve according to Navier--Stokes equations. Notice that the theorem is only useful when the inequality is violated. However, checking whether (\ref{eq section 2 026}) is violated or not for every sphere in the velocity field region at every time could be a huge computational effort. 
To reduce this computational effort, we would like to avoid working out the $\partial_i \partial_j p$ term inside integrand but, at the same time,  we would like to preserve the Poisson equation for pressure and to avoid vorticity equation incompatibility. Tentatively, we could take the quantities inside brackets of the first term in the integrand of (\ref{eq section 2 026}) as antisymmetric matrix components. Namely, we could take $A_{ij}=-A_{ji}$ and
\begin{eqnarray}
A_{ij}=2\partial_ju_k\partial_ku_i+\partial_i\partial_j p-\nu\partial_k\partial_k \partial_i u_j.
\label{eq section 3 0001}
\end{eqnarray}
In this case, the second time derivative of surface area would be positive for all surfaces, not only for spheres. But the trace of every antisymmetric matrix is null. Computing the trace of this antisymmetric matrix and taking into account the pressure Poisson equation, we obtain that 
\begin{eqnarray}
-\partial_i u_j \partial_j u_i=\partial_i\partial_i p=0
\label{eq section 3 0002}
\end{eqnarray}
for incompressible Navier--Stokes equations. In this way, the pressure is a harmonic function. However, in the next section we will address no-slip boundary conditions on velocity fields and this type of restrictions would produce no effect. Moreover, instead of reduce the computational task of (\ref{eq section 2 026}), by taking the antisymmetric matrix (\ref{eq section 3 0001}), we can circumvent it in other manner. The alternative to obtain more velocity field properties is the following. In addition to Navier--Stokes equations
\begin{eqnarray}
\partial_tu_i+u_k\partial_ku_i=-\partial_ip+\nu \partial_k\partial_ku_i\nonumber\\
\partial_ku_k=0,
\label{eq section 3 001}
\end{eqnarray}
we could propose the heuristic relation between velocity and pressure given by
\begin{eqnarray}
\frac{1}{2}\partial_iu_k\partial_ku_j +\frac{1}{2}\partial_ju_k\partial_ku_i+\left(\Tr(M)\delta_{ij}-3M_{ij}\right)+\partial_i\partial_j p=0,
\label{eq section 3 002}  
\end{eqnarray}
where $M$ is a symmetric matrix with trace $\Tr(M)=M_{ij}\delta_{ij}$ and components $M_{ij}=M_{ji}$ that are arbitrary functions of $\bi{x}$ and $t$. This relation is inspired in the quantities inside brackets of the first term in the integrand of (\ref{eq section 2 026}). Now, relation (\ref{eq section 3 002}) is compatible with the Poisson equation for pressure $\partial_iu_j\partial_ju_i=-\partial_k\partial_kp$ and vorticity equation. This matrix adds several degrees of freedom for this pressure equation since we are free to choose $M_{ij}$. There are six degrees of freedom corresponding to six arbitrary ways to choose $M_{ij}$. However, this relation also imposes six independent equations to velocity and pressure relationship while there are four unknowns. This could make the system of equations inconsistent. To prevent the system from being overdetermined, it is necessary that we add an unknown term to the momentum equation in (\ref{eq section 3 001}). In following section it moves ahead for fitting the parameters and unknown functions.

\section{Dirichlet boundary condition on Restricted Navier-Stokes equations.}

Now that we have found a system of partial differential equations, we will focus on the Dirichlet problem. A problem in the Navier--Stokes equations (\ref{eq section 3 001}) is to manage the dissipative term
\begin{eqnarray}
\nu\int_{\Omega_t}u_i\partial_k\partial_ku_id^3x\label{eq section 4 001}
\end{eqnarray}
in the energy equation. Where there are the boundary conditions, it can be worked out as follows. As usual \cite{RefB2}, \cite{RefB101}; when computing the integral of the square of vorticity $\omega_i=\epsilon_{ijk}\partial_j u_k$ taken over the volume under no-slip condition (i.e., $u_i=0$) on the boundary $\partial \Omega_t$, the time derivative of the kinetic energy is, 
\begin{eqnarray}
\frac{1}{2}\frac{d}{dt}\int_{\Omega_t}u_iu_id^3x=\nu\int_{\Omega_t} u_i \partial_k\partial_k u_i d^3x =-\nu\int_{\Omega_t} \omega_i \omega_i d^3x \leq 0.
\label{eq section 4 0002}  
\end{eqnarray}
 
Although the next theorem will give us a similar result, it lets us to introduce other one. This last theorem is the main issue of this work. The first theorem comprises Navier--Stokes equations, the additional relation between velocity and pressure, along with Dirichlet boundary conditions. The second theorem uses this result to delimiting the kinetic energy of usual Navier--Stokes equations with no-slip boundary.

\subsubsection*{\bf{Theorem 4.}}Let $\mathbb{R}^3$ be the Euclidean space. Let $u_i,p,f_i$ be class $C^3$ differentiable functions on $\mathbb{R}^3\times [t_0,\infty)$ as defined in the previous section and $i\in\{1,2,3\}$. Take $\lambda>0,\nu>0$  constants. The volume $\Omega_t$ is compact at time $t$ with smooth boundary $\Sigma_t$. Suppose that $u_i,p,f_i$ satisfy
\begin{eqnarray}
\frac{Du_i}{Dt}=\nu \partial_k\partial_k u_i -\partial_i p +f_i\label{eq section 4 0011}\\
\partial_i u_i=0 \label{eq section 4 0012}\\
N_{ij}=\frac{1}{2}\left(\partial_iu_k\partial_ku_j +\partial_ju_k\partial_ku_i\right)-M_{ij}+\partial_i\partial_j p \label{eq section 4 0013}\\
M_{ij}=\frac{\nu}{2}\partial_k\partial_k(\partial_j u_i+\partial_i u_j) +\frac{\lambda}{4}(\partial_j u_i+\partial_i u_j)\label{eq section 4 0015}\\
2N_{ij}-\partial_i f_j -\partial_j f_i=0\label{eq section 4 0017}
\end{eqnarray}
in the volume $\Omega_t$, and  $u_i$ satisfies
\begin{eqnarray}
u_i=0
\label{eq section 4 002}  
\end{eqnarray}
in the surface $\Sigma_t$. Then, it also satisfies
\begin{eqnarray}
\int_{\Omega_t}u_iu_id^3x\leq \left[\sqrt{K_2}+\sqrt{K_3\left(t\right)}\right]^2
\label{eq section 4 003}  
\end{eqnarray}
where
\begin{eqnarray}
K_2=\int_{\Omega}u^0_iu^0_id^3x
\label{eq section 4 004}
\end{eqnarray}
is constant and
\begin{eqnarray}
K_3\left(t\right)=\frac{1}{4}\int_{t_0}^t\int_{\Omega_\tau}f_if_id^3xd \tau.
\label{eq section 4 005}  
\end{eqnarray}
Here $u^0_i$ and $\Omega$ are $u_i$ and $\Omega_t$ at time $t_0$, respectively.
\subsubsection*{Proof.} Notice that we can obtain
\begin{eqnarray}
\frac{1}{2}\frac{D\left(u_iu_i\right)}{Dt}=\nu u_i \partial_k\partial_k u_i -\partial_i\left( u_ip \right) +u_if_i
\label{eq section 4 006}  
\end{eqnarray}
multiplying (\ref{eq section 4 0011}) by $u_i$. Integrating in $\Omega_t$, using Reynolds transport theorem and (\ref{eq section 4 0012}), we have
\begin{eqnarray}
\frac{1}{2}\frac{d}{dt}\int_{\Omega_t}u_iu_id^3x=\nu\int_{\Omega_t} u_i \partial_k\partial_k u_i d^3x+\int_{\Omega_t} u_if_id^3x-\int_{\Sigma_t}\left( u_ip \right)n_id^{3}x.
\label{eq section 4 007}  
\end{eqnarray}
Notice that $\Sigma_t=\Sigma$ since $u_i=0$ in the boundary. This causes that the last term of right hand side disappears. The first term in the right hand side can be worked out as follows. Taking derivative of momentum equation (\ref{eq section 4 0011}) and substitution of (\ref{eq section 4 0013}) in the resulting one gives
\begin{eqnarray}
\frac{D}{Dt}(\partial_ju_i)=\frac{1}{2}\partial_iu_k\partial_ku_j
-\frac{1}{2}\partial_ju_k\partial_ku_i-N_{ij}-M_{ij}\nonumber\\
+\partial_jf_i+
\nu\partial_k\partial_k\partial_ju_i.\label{eq section 4 008}
\end{eqnarray}
The product of this equation by $(\partial_ju_i+\partial_iu_j)$ is
\begin{eqnarray}
\frac{1}{4}\frac{D}{Dt}[(\partial_iu_j+\partial_ju_i)(\partial_iu_j+\partial_ju_i)]=\partial_iu_j\left[\nu\partial_k\partial_k(\partial_iu_j+\partial_ju_i)\right]\nonumber\\
-\left(2N_{ij}-\partial_if_j-\partial_jf_i\right)\partial_j u_i
-2M_{ij}\partial_iu_j.
\label{eq section 4 009}  
\end{eqnarray}
When we choose the symmetric matrix $M$ with the components $M_{ij}$ given by (\ref{eq section 4 0015}), replace them in (\ref{eq section 4 009}), using (\ref{eq section 4 0017}) and (\ref{eq section 4 0012}), it gives
\begin{eqnarray}
\frac{D}{Dt}[(\partial_iu_j+\partial_ju_i)(\partial_iu_j+\partial_ju_i)]=-\lambda (\partial_iu_j+\partial_ju_i)(\partial_iu_j+\partial_ju_i).
\label{eq section 4 010}  
\end{eqnarray}
We now integrate (\ref{eq section 4 010}) in $\Omega_t$. Using Reynolds transport theorem, we obtain
\begin{eqnarray}
\frac{d}{dt}\int_{\Omega_t}(\partial_iu_j+\partial_ju_i)(\partial_iu_j+\partial_ju_i)d^3x\nonumber\\
=-\lambda\int_{\Omega_t}(\partial_iu_j+\partial_ju_i)(\partial_iu_j+\partial_ju_i)d^3x.
\label{eq section 4 0101} 
\end{eqnarray}
The time integral on interval $[t_0, t)$ gives us
\begin{eqnarray}
\int_{\Omega_{t}}(\partial_j u_i+\partial_i u_j)(\partial_j u_i+\partial_i u_j) d^3 x
=K_4 e^{-\lambda\left(t-t_0\right)},
\label{eq section 4 011}  
\end{eqnarray}
where
\begin{eqnarray}
K_4=\int_{\Omega_{t}}(\partial_j u^0_i+\partial_i u^0_j)(\partial_j u^0_i+\partial_i u^0_j) d^3 x.
\label{eq section 4 0111}  
\end{eqnarray}
Rearranging terms, we have
\begin{eqnarray}
2 \int_{\Omega_{t}}\partial_j\left[\left(\partial_j u_i+\partial_i u_j\right) u_i\right]d^3 x-2\int_{\Omega_{t}}u_i\partial_j\partial_j u_id^3 x
=K_4 e^{-\lambda\left(t-t_0\right)} .
\label{eq section 4 012}  
\end{eqnarray}
Applying Gauss theorem and boundary condition to this last formula gives us
\begin{eqnarray}
\int_{\Omega_{t}}u_i\partial_j\partial_j u_id^3 x
=-\frac{K_4}{2} e^{-\lambda\left(t-t_0\right)}. 
\label{eq section 4 013}  
\end{eqnarray}
Substitution of this relation on (\ref{eq section 4 007}) gives rise to
\begin{eqnarray}
\frac{1}{2}\frac{d}{dt}\int_{\Omega_t}u_iu_id^3x=-\frac{\nu K_4}{2}e^{-\lambda(t-t_0)}+\int_{\Omega_t}f_iu_id^3x.
\label{eq section 4 014}  
\end{eqnarray}
The first term of right hand side in this last formula is negative and the second one can be approximated by Cauchy-Schwartz inequality. So, (\ref{eq section 4 014}) can be approximated by
\begin{eqnarray}
\frac{1}{2}\frac{d}{dt}\int_{\Omega_t}u_iu_id^3x\leq\sqrt{\int_{\Omega_t}f_if_id^3x}\sqrt{\int_{\Omega_t}u_iu_id^3x}.
\label{eq section 4 014.1}  
\end{eqnarray}
Notice that this is a time differential inequality of Bihari--LaSalle's type (see \cite{RefJ11}, \cite{RefJ111})
\begin{equation}
\frac{d F(t)}{dt}\leq C(t)\sqrt{F(t)}.
\label{eq theorem 2 008} 
\end{equation}
Dividing (\ref{eq theorem 2 008}) by \(\sqrt{F(t)}\), and integrating in time variable the resulting relation is
\begin{equation}
\sqrt{F(t)}\leq \sqrt{F(t_0)}+\int_{t_0}^t\frac{C(\tau)}{2}d\tau.
\label{eq theorem 2 009}
\end{equation} 
Then, we conclude that
\begin{eqnarray}
\int_{\Omega_t}u_iu_id^3x\leq \left[\sqrt{K_2}+\sqrt{K_3(t)}\right]^2
\label{eq section 4 016}.
\end{eqnarray} 
\hfill\ensuremath{\square}

A consequence of Theorem 4 is that when we choose the arbitrary functions $f_i$ such as they have finite $L^2$ norm in $\mathbb{R}^3\times[t_0, \infty)$, then
\begin{eqnarray}
\int_{\Omega_t}u_iu_id^3x<\infty 
\label{eq section 4 0162}  
\end{eqnarray}
for all $t\in\left[t_0,\infty\right)$. So, the kinetic energy of the system does not blow up under the considered conditions. Although holding bounded this energy is a long standing problem in Navier--Stokes equations, relation (\ref{eq section 4 003}) is an inequality that does not provides all the system information. However, we can consider a sequence of functions $\left\{f^{(n)}_i\right\}_{n\in \mathbb{N}}$ such as $f^{(n)}$ replace $f_i$ in (\ref{eq section 4 0011})-(\ref{eq section 4 0017}). When this sequence holds $f^{(n)}_i\rightarrow 0$ when $n\rightarrow \infty$, we can compute time integral of (\ref{eq section 4 014}) as
\begin{eqnarray}
\int_{\Omega_t}u_iu_id^3x\rightarrow K_2+\frac{\nu K_4}{\lambda}\left[e^{-\lambda\left(t-t_0\right)}-1\right].
\label{eq section 4 0163}  
\end{eqnarray}
Then, the highest value of the energy depends only on initial conditions. Moreover, it is necessary to take $K_2\geq \frac{\nu K_4}{\lambda}$ to avoid contradictions at time $t\rightarrow \infty$ in the sequence limit. Then, the chance of choice the additional relation between velocity and pressure makes it possible to bound highest growth of the energy that is put into play by the system. However, if we only start with Navier--Stokes equations and Dirichlet boundary conditions, we have several options to choose the symmetric matrix $M$ and, hence, several solutions. This can be enunciated as a theorem.
\subsubsection*{\bf{Theorem 5.}} Let $\mathbb{R}^3$ be the Euclidean space. Let $u_i,p$ be class $C^3$ differentiable functions on $\mathbb{R}^3\times [t_0,\infty)$ as defined above and $i\in\{1,2,3\}$. Take $\nu>0$ constant. And $\Omega_t$ is a compact volume at time $t$ with piecewise smooth boundary $\Sigma_t$. Suppose that $u_i,p$ satisfy
\begin{eqnarray}
\frac{Du_i}{Dt}=\nu \partial_k\partial_k u_i -\partial_i p\label{eq section 4 0161}\\
\partial_i u_i=0 \label{eq section 4 00162}
\end{eqnarray}
in the volume $\Omega_t$, and  $u_i$ satisfies
\begin{eqnarray}
u_i=0
\label{eq section 4 017}  
\end{eqnarray}
in the surface $\Sigma_t$. Here $u^0_i$ and $\Omega$ are $u_i$ and $\Omega_t$  at time $t_0$, respectively. Then, the solution 
\begin{eqnarray}
\int_{\Omega_t}u_iu_id^3x<\infty
\end{eqnarray}
is not unique.
\subsubsection*{Proof.}Suppose that
\begin{eqnarray}
\int_{\Omega}\left(\partial_ju^0_i+\partial_iu^0_j\right)\left(\partial_ju^0_i+\partial_iu^0_j\right)d^3x\neq 0.
\label{eq section 4 018}  
\end{eqnarray}
Let be $a$ an element of a set of indices $J$. We can use Theorem 4 for constants $\lambda^{(a)}$ and function sequence $\left\{f^{(n)}_i\right\}_{n\in \mathbb{N}}$ as before, in such a way that $f^{(n)}_i\rightarrow 0$. Then, we have
\begin{eqnarray}
\int_{\Omega_t}u_iu_id^3x\rightarrow K_2+\frac{\nu K_4}{\lambda^{(a)}}\left[e^{-\lambda^{(a)}\left(t-t_0\right)}-1\right]
\label{eq section 4 019}  
\end{eqnarray}
for all $a\in J$ when $n\rightarrow \infty$. But we can impose $\lambda^{(a)}\neq\lambda^{(b)}$ for $a,b\in J$, $a\neq b$. But the integral of functions in $L^2$  can not converge to several values at the same time, since it violates bounded convergence theorem of Lebesgue integral.

In the case that
\begin{eqnarray}
\int_{\Omega}\left(\partial_ju^0_i+\partial_iu^0_j\right)\left(\partial_ju^0_i+\partial_iu^0_j\right)d^3x= 0
\label{eq section 4 021}  
\end{eqnarray}
we have
\begin{eqnarray}
\int_{\Omega_t}\left(\partial_ju_i+\partial_iu_j\right)\left(\partial_ju_i+\partial_iu_j\right)d^3x= 0
\label{eq section 4 022}  
\end{eqnarray}
for all $t\in[t_0, \infty)$. So $(\partial_ju_i+\partial_iu_j)=0$ and, hence, $\partial_j\partial_ju_i=0$ for $\bi{x}\in \Omega_t$. But $u_i=0$ in $\Omega_t$ if $u_i$ is a harmonic function with $u_i=0$ in $\Sigma_t$.\hfill\ensuremath{\square}

One can wonder whether there are several $n\in \mathbb{N}$ for which $f^{(n)}_i$ gives the same result or not. We can address this issue considering, instead of the external field function sequence, only one field function with a constant going to zero. For example, since the external force field components $f_i$ are arbitrary, we can impose $f_i =-\frac{1}{2} \kappa u_i$ in (\ref{eq section 4 0011})-(\ref{eq section 4 0017}). This constant holds that $\kappa>0$ , $\kappa\rightarrow 0$ and $\kappa\neq\lambda$. Then, the equation (\ref{eq section 4 014}) becomes
an inhomogeneous linear differential equation of kinetic energy which exact solution is
\begin{eqnarray}
\int_{\Omega_t}u_iu_id^3x= K_2+\frac{\nu K_4}{\lambda-\kappa}\left[e^{-\lambda\left(t-t_0\right)}-e^{-\kappa\left(t-t_0\right)}\right].
\label{eq section 4 01621}  
\end{eqnarray}
Notice that here the kinetic energy reaches a minimum at time 
\begin{eqnarray}
t_{min}=t_0+\frac{1}{\lambda-\kappa}\ln \left(\frac{ \lambda}{\kappa}\right).
\label{eq section 4 01622}
\end{eqnarray}
Hence, to avoid incompatibilities, it is necessary that kinetic energy would never become negative, even at the time in which it is minimum. In this way,  $\kappa$ and $\lambda$ are related by
\begin{eqnarray}
K_2\geq -\frac{\nu K_4}{\lambda-\kappa}\left[\left(\frac{\kappa}{\lambda}\right)^{\frac{\lambda}{\lambda-\kappa}}-\left(\frac{\kappa}{\lambda}\right)^{\frac{\kappa}{\lambda-\kappa}}\right].
\label{eq section 4 01623}  
\end{eqnarray}

So we can conclude that Navier--Stokes equations with the Dirichlet boundary conditions given above have not unique solution. Moreover, the solution for restricted Navier--Stokes equations has, in the worst case, $\sim O\left(K_3(t)\right)$ growing with time. In the best case, when the sequence of functions make $f^{(n)}_i$ going to zero, the solution decays exponentially with time. This stability for $t\rightarrow \infty$ of (\ref{eq section 4 0163}) is in good agreement with \cite{RefJ5.1}, \cite{RefJ13} and fluid phenomena observed in experiments \cite{RefJ12}. Moreover, the problem for computing whether the fluid field holds (\ref{eq section 2 026}) or not can be circumvented in this way.
\section{Conclusion}
This paper has shown the usefulness of considering the movement of the surface of a volume dragged by a velocity field. When the surfaces are spheres, it is needed to work out (\ref{eq section 2 016.3}), the second derivative of surface area with respect to time. It allows the chance to find the  relation (\ref{eq section 2 026}) to avoid unrealistic velocity fields that do not evolve according to Navier--Stokes equations. However, to check this relation for every sphere in the considered domain of the field supposes a hard computational task. This difficulty can be overcome by taking another strategy. We can make the ansatz (\ref{eq section 3 002}), with several degrees of freedom, in such a way that there is no contradiction with pressure Poisson equation, and then, we can proceed by tuning such degrees of freedom. We are free to choose a symmetric matrix which, under suitable boundary conditions, gives rise to bound essential magnitudes. When Dirichlet no-slip conditions are applied on the boundary of the domain of this dynamical system, the kinetic energy on considered volume increases at most $\sim O\left(K_3(t)\right)$  with time as (\ref{eq section 4 003}). But since there is an arbitrary choice of $\lambda$, only considering incompressible Navier--Stokes equations with those boundary conditions, we conclude that we can obtain several solutions simultaneously, as viewed in Theorem 5. Moreover, taking the external force components as $f_i=-\frac{1}{2}\kappa u_i$, for any $\kappa>0$ holding $\kappa\neq \lambda$ and (\ref{eq section 4 01623}), we can compute the kinetic energy (\ref{eq section 4 01621}) of the system (\ref{eq section 4 0011})-(\ref{eq section 4 002}) without approximation. In this way, it is not necessary to consider approximating artefacts, such as the artificial compressibility, in numerical methods. Moreover, it is remarkable that those solutions are not weak, but classical, since they are not class $C^\infty$ functions, but $C^3$ functions.
\section{Acknowledgements}
I thank M.Casado and M.J.Campos for moral support and understanding.

\end{document}